\documentclass{JINST}

\title {\boldmath Measurement of the $K_L$ nuclear interaction
length in the NaI(Tl) calorimeter }

\author{
M.~N.~Achasov$^a$$^b$, 
K.~I.~Beloborodov$^a$$^b$\thanks{Corresponding author.}, 
A.~V.~Berdyugin$^a$$^b$, 
A.~G.~Bogdanchikov$^a$, 
A.~V.~Vasiljev$^a$$^b$, 
V.~B.~Golubev$^a$$^b$, 
T.~V.~Dimova$^a$$^b$, 
V.~P.~Druzhinin$^a$$^b$, 
A.~A.~Korol$^a$$^b$, 
S.~V.~Koshuba$^a$, 
E.~V.~Pakhtusova$^a$, 
S.~I.~Serednyakov$^a$$^b$, 
Z.~K.~Silagadze$^a$$^b$ and
Yu.~V.~Usov$^a$$^b$\\
\llap{$^a$}Budker Institute of Nuclear Physics,                                                                                            
  630090 Novosibirsk, Russia\\                                                                                                       
\llap{$^b$}Novosibirsk State University,                                                                                            
  630090 Novosibirsk, Russia\\                                                                                                       
  E-mail: \email{K.I.Beloborodov@inp.nsk.su}}

\abstract{
In the study of the reaction $e^+e^-\to K_{S}K_{L}$ at the VEPP-2M 
$e^+e^-$  collider with the SND detector the nuclear interaction
length of $K_{L}$ meson in NaI(Tl) has been measured. Its value is found 
to be 30--50 cm in the $K_{L}$ momentum range 0.11--0.48 GeV/$c$.
The results are compared with the values used in the 
simulation programs GEANT4 and UNIMOD.
}


\begin{document}

\section{Introduction}
In the analysis of experimental data in the high-energy physics
the Monte-Carlo (MC) simulation is widely used. There are known 
simulation packages,  such as  UNIMOD~\cite{UNIMOD}, GEANT3~\cite{geant3},
and GEANT4~\cite{GEANT}, which successfully describe the physical processes
of particles passage through the detector material. 
Nevertheless, in some cases, the accuracy of the 
theoretical models and/or the experimental data used in simulation 
may be not sufficient. Therefore, it is worthwhile to verify the simulation
using new experimental data.

In $e^+e^-$ annihilation into hadrons at the center-of-mass (c.m.) energies
above 1 GeV the neutral kaons are intensively produced, e.g. in the 
process 
\begin{equation}
e^+e^-\to K_SK_L.
\label{kskl}
\end{equation}
Whereas $K_S$ meson decays rapidly via $K_S\to\pi\pi$ channel,
the neutral $K_L$ meson, due to its much greater lifetime,  passes through
detector material and is absorbed due to the nuclear interaction.
Since kaons in the process (\ref{kskl}) are 
producing in pairs, the $K_L$ meson can be tagged  by the  
recoiled $K_S$ meson. This greatly facilitates the study of $K_L$ 
interaction with the detector material.

In this paper we use kaons from the reaction (\ref{kskl}) to
measure the $K_L$ nuclear interaction length in 
NaI(Tl) crystals in the momentum range from 0.11 to 0.48 GeV/$c$.

\section{Experiment}
The  experiment on the study the $e^+e^-\to K_{S}K_{L}$ process
was carried out at the VEPP-2M~\cite{V2M} $e^+e^-$  collider with 
the SND detector~\cite{SND}
in the c.m. energy range  $\sqrt{s}=2E=$0.98--1.38 GeV.
The SND is a general purpose non-magnetic detector
for low energy $e^+e^-$-colliders. In the center of the detector around 
the collider beam pipe a tracking system consisting of two drift chambers
is installed. The main part of the SND is a three-layer spherical
electromagnetic calorimeter based on 1640 NaI(Tl) crystals  with the total
weight 3.6 ton. The calorimeter covers the polar angle range 
$18^\circ<\theta<162^\circ$. The crystal angular dimension  
is $\Delta\theta=\Delta\phi=9^\circ$. The radial positions of the
calorimeter layers are shown in Fig.~\ref{fig3}. 
The total calorimeter thickness for particles originating from the 
interaction region is 34.7 cm (13.4$X_0$), where $X_0$ is the 
radiation length equal to 2.6 cm for NaI(Tl). 
Pairs of crystals of the two inner layers with thickness of 2.9 and 4.8$X_0$
are sealed in a common thin (0.1 mm) aluminum container. In the gap between
the second and third calorimeter layers an aluminum supporting hemisphere is 
located.
The scintillation light signals from the crystals are detected by vacuum 
phototriodes. Outside the calorimeter an iron absorber and 
a muon detector are placed. 

In this paper the data collected 
in the energy interval $\sqrt{s}=$1.04--1.38 GeV with an integrated
luminosity  of~9.2 pb$^{-1}$ are analyzed. These data were used 
previously to measure the $e^+e^-\to K_{S}K_{L}$ cross section~\cite{eeksl}.
In addition, the data recorded at the maximum of the $\phi(1020)$ resonance
with an integrated luminosity of~0.5 pb$^{-1}$ are used.

The Monte-Carlo event generator for the $e^+e^-\to K_{S}K_{L}$ reaction 
used in this analysis includes radiative corrections~\cite{kuraev}. 
In particular, an extra photon emitted by initial electrons is generated 
with the angular distribution modelled according to Ref.~\cite{BM}.
The $e^+e^-\to K_{S}K_{L}$ Born cross section needed to calculate
the radiative corrections is taken from Ref.~\cite{eeksl}.

The response of the SND detector is simulated using the UNIMOD 
package~\cite{UNIMOD}. This package developed in BINP (Novosibirsk) in 
eighteens is functionally similar to the GEANT3~\cite{geant3} simulation
program. Special attention in UNIMOD is given to simulation of pion and kaon 
low-energy nuclear interactions. Nuclear cross sections are 
calculated with the SCATTER program~\cite{SCATTER}, while the
nuclear reactions are simulated using the NUCRIN model~\cite{NUCRIN}.
The simulation takes into account the variation of the detector and 
accelerator conditions (dead electronic channels, size of the
interaction region, etc.) and beam-induced background photons and 
charged particles overlapping events of interest.

\section{Selection of $e^+e^-\to K_SK_L$ events}
\begin{figure}
\centering
\includegraphics[width=0.6\textwidth]{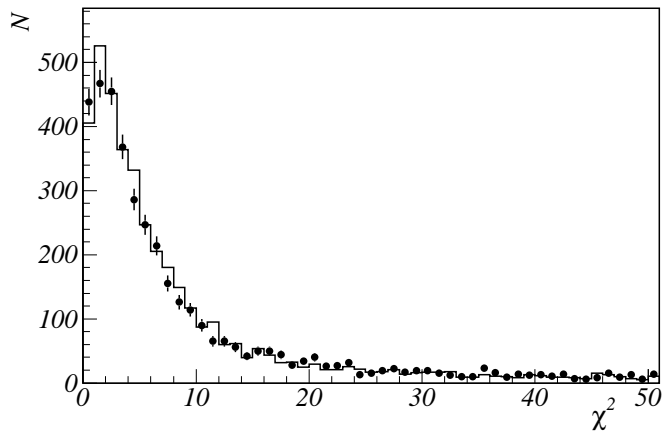}
\caption{\label{fig1}
The distribution of $\chi^2$ of the kinematic fit to
the $K_S \to \pi^0\pi^0 \to 4\gamma$ hypothesis 
for $K_S$ candidates in data (points with error bars)
and $e^+e^-\to K_SK_L(\gamma)$ simulation (histogram)
at $\sqrt{s}=1.04-1.10$ GeV.}
\end{figure}
   The $K_S$ mesons from the process $e^+e^-\to K_SK_L$ 
decay inside the SND tracking system.    
The decay mode $K_S\to\pi^0\pi^0\to 4\gamma$ is chosen for $K_S$ 
reconstruction. Therefore we select events containing at least four
photons and no charged particles.
To eliminate the cosmic ray background, the veto from muon detector and 
the requirement that the fired calorimeter crystals do not lie along a 
straight line are used.
\begin{figure}
\centering
\includegraphics[width=0.6\textwidth]{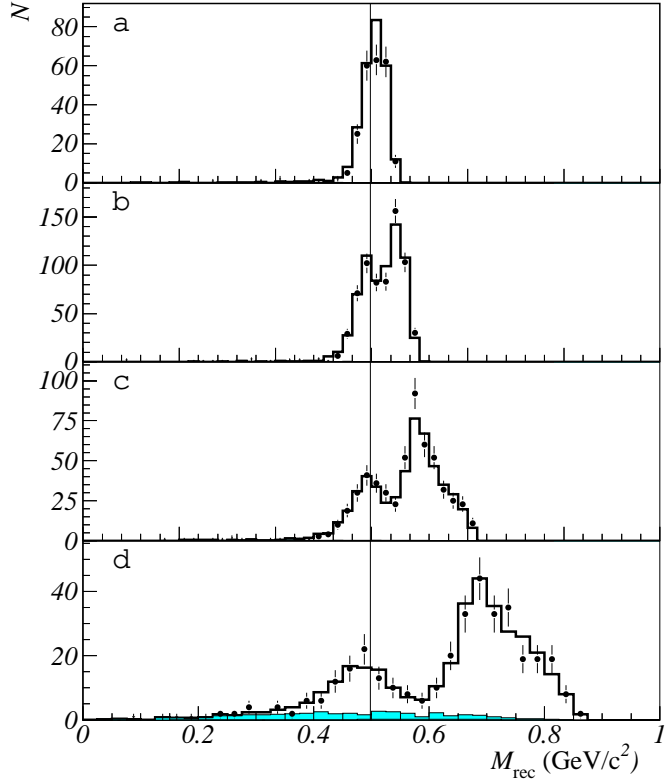}
\caption{\label{fig2}
The $M_{\rm rec}$ spectrum for selected $e^+e^-\to K_SK_L$ candidates
in four energy intervals: (a) $\sqrt{s}=1.04-1.05$ GeV,
(b) $\sqrt{s}=1.06-1.09$ GeV, (c) $\sqrt{s}=1.10-1.20$ GeV,  and
(d) $\sqrt{s}=1.20-1.38$ GeV.
The shaded histogram shows the $e^+e^-\to\omega\pi^0$
background. The vertical line indicates $K_L$ mass.}
\end{figure}

We combine four photons in an event to create a $K_S$ candidate.
Photons included into the $K_S$ candidate must be in the polar angle 
range $36^\circ<\theta_{\gamma}<144^\circ$ and have the transverse 
energy profile in the calorimeter corresponding to the profile  
from single electromagnetic shower~\cite{xinm}. The four photons are
kinematically fitted to the $K_S \to \pi^0\pi^0 \to 4\gamma$ hypothesis.
The distribution of $\chi^2$ of the kinematic fit is shown in Fig.~\ref{fig1}
for data and simulated $e^+e^-\to K_SK_L$ events from the energy
region $\sqrt{s}=1.04-1.10$ GeV, in which the background contribution
is small. If more than one $K_S$ candidate is found in an event, 
the candidate with the lowest $\chi^2$ value is chosen. 
For further analysis, events with $\chi^2<25$ are selected.

The dominant background process is
$e^+e^-\to\omega\pi^0\to\pi^0\pi^0\gamma$.
To suppress this background the kinematic fit to the
$e^+e^-\to\pi^0\pi^0\gamma$ hypothesis is
performed for events containing five or more photons.
Events with $\chi^2_{\pi^0\pi^0\gamma}<60$
are rejected.

The measured $K_S$ parameters are used to calculate the mass recoiling
%
against the $K_S$: $M_{\rm rec}=\sqrt{(2E-E_{K_S})^2-p_{K_S}^2}$, 
where $E_{K_S}$ and $p_{K_S}$ are the $K_S$ energy and momentum, respectively.
The $M_{\rm rec}$ spectra for different 
energy intervals are shown in Fig.~\ref{fig2}. 
The two peaks in the spectra for energies $\sqrt{s}> 1.06$ GeV
correspond to the reactions $e^+e^-\to K_SK_L$ and $e^+e^-\to K_SK_L\gamma$.
The latter reaction is dominated by the radiative return to the $\phi(1020)$
resonance, i.e. the process $e^+e^-\to \phi\gamma$.
In the following analysis we use events with the recoil mass close to
$K_L$ mass ($0.40 < M_{\rm rec} < 0.55$ GeV/$c^2$), which come
from the $e^+e^-\to K_SK_L$ reaction.

The distribution of selected events over energy intervals
is given in Table~\ref{tab1}. In the energy range 
above the $\phi(1020)$ resonance  ($\sqrt{s}\geq 1.04$ GeV) 
2606 events are selected. The background sources were studied
in detailed in our earlier work~\cite{eeksl}. The expected number of 
background events at $\sqrt{s}\geq 1.04$ GeV is $73\pm7$ 
(36 from the beam background, 25 from $e^+e^-\to \eta\gamma$, 
and 12 from $e^+e^-\to \omega\pi^0$). 

\section{Extraction of the $K_L$ nuclear interaction length}
The $K_L$ decay length $\lambda_{\rm dec}$ varies from 3.4 m
at $E=0.51$ GeV to 15.2 m at $E=0.7$ GeV and far
exceeds the calorimeter outer radius, about 0.7 m. 
The nuclear interaction length $\lambda_{\rm int}$ is close
to the thickness of the calorimeter sensitive volume $L_{\rm cal}=34.7$ cm.
Therefore, a significant part of $K_L$ mesons undergoes a
nuclear interaction in the calorimeter.
Charged particles and photons produced in the interaction
give an energy deposition in NaI(Tl) crystals near the
$K_L$ path. 
Events selected using the criteria described above
are divided into two classes, with four and with five or
more photons. The latter class is dominated by
events, in which the $K_L$ mesons either decay or interact 
inside the detector volume. 
\begin{figure}
\centering
\includegraphics[width=0.6\textwidth]{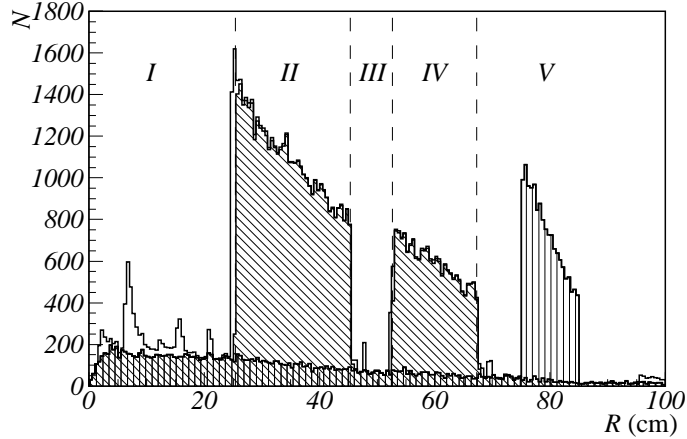}
\caption{\label{fig3}
The simulated distribution of the radius of the disappearance point 
(due to decay or nuclear interaction) for $K_L$ mesons with $E=0.51$ GeV.
The cross-hatched histogram 
shows the contribution of $K_L$ decays.
The diagonally hatched areas in the zones II and IV show
the contribution of nuclear interaction in NaI(Tl) crystals.
The unhatched areas in the zones I, III and V represent the contribution 
of nuclear interaction in the detector material before the 
calorimeter, between the calorimeter layers, and
after the calorimeter, respectively. The interaction in the iron absorber
is shown by the histogram with vertical hatching. The deviation
of the distribution for $K_L$ decays from the exponential
near zero is due to the finite size of the beam interaction
region along the collider beam axis 
(radius is measured from the geometric center of the detector).}
\end{figure}

Figure~\ref{fig3} shows the simulated radial distribution of the
disappearance point (due to decay or nuclear interaction) for $K_L$ mesons
with $E=0.51$ GeV. It is seen that
the main mechanism of the $K_L$ disappearance inside the detector
(zones I--IV in Fig.~\ref{fig3}) is a nuclear 
interaction in NaI(Tl). From the experiment we extract  
the ratio of the number of events with five or more photons
($N_{5\gamma}-N_{5\gamma,{\rm bkg}}$)
to the total number of events ($N-N_{\rm bkg}=
N_{5\gamma}+N_{4\gamma}-N_{5\gamma,{\rm bkg}}-N_{4\gamma,{\rm bkg}}$), where
$N_{4\gamma}$ ($N_{5\gamma}$) is the number of selected events with four 
(five or more) photons, and $N_{4\gamma,{\rm bkg}}$ 
($N_{5\gamma,{\rm bkg}}$) is the estimated number of background 
events with four (five ore more) photons.

The number of events with five or more photons can be calculated as
follows:
\begin{eqnarray}
N_{5\gamma}^{\rm calc}&=&N_0\sum_{i={\rm I}}^{\rm V}r_i, \nonumber\\
r_{\rm I}&=&w_{\rm I}\varepsilon^{5\gamma}_{\rm I}, \nonumber\\
r_{\rm II}&=&w_{\rm II}(1-w_{\rm I})\varepsilon^{5\gamma}_{\rm II}, \nonumber\\
r_{\rm III}&=&w_{\rm III}(1-w_{\rm I})(1-w_{\rm II})\varepsilon^{5\gamma}_{\rm III}, \nonumber\\
r_{\rm IV}&=&w_{\rm IV}(1-w_{\rm I})(1-w_{\rm II})(1-w_{\rm III})\varepsilon^{5\gamma}_{\rm IV}, \nonumber\\
r_{\rm V}&=&(1-w_{\rm I})(1-w_{\rm II})(1-w_{\rm III})(1-w_{\rm IV})\varepsilon^{5\gamma}_{\rm V}, \nonumber
\label{eq1}
\end{eqnarray}
where $N_0$ is the number of $e^+e^-\to K_SK_L$ data events.
The probabilities of the $K_L$ disappearance in the NaI(Tl)
(zones II and IV in Fig.~\ref{fig3}) are
equal to $w_{\rm II,IV}=1-\exp(-L_{\rm II,IV}/\lambda)$, where
$1/\lambda=1/\lambda_{\rm int}+1/\lambda_{\rm dec}$,
$L_{II}$ is the total thickness of the 1-st and 2-nd calorimeter layers,
and $L_{IV}$ is the  thickness of the third layer.
The $\lambda_{\rm dec}$ is calculated as an average over detected events
from the energy interval under study.

%
The probabilities of $K_L$ decay and nuclear interaction before the
calorimeter $w_{\rm I}$ and between the calorimeter layers
$w_{\rm III}$,  as well as
the probabilities for an event to have five or more photons due to
the disappearance in the $i$-th detector zone $\varepsilon^{5\gamma}_i$
are obtained using MC simulation. The zone V in Fig.~\ref{fig3} contains 
$K_L$ decays or interactions outside the calorimeter.
Even in the absence of the $K_L$ signal an event can be selected  
to the five-photon class, for example, because of the splitting
electromagnetic shower of one of the photons from $K_S$ decay.

The same formulae with the replacement of $\varepsilon^{5\gamma}_i$
by $\varepsilon^{4\gamma}_i$ are used to calculate the number of four-photon
events. From the comparison of the predicted ratio
$N_{5\gamma}^{\rm calc}/(N_{5\gamma}^{\rm calc}+N_{4\gamma}^{\rm calc})$ 
with the ratio obtained in data,
the initial value of $\lambda_{\rm int}$ is obtained. The initial
value is then corrected to take into account the following effects.

\begin{table}
\centering
\caption{\label{tab1} The energy interval
($\sqrt{s}$), integrated luminosity ($IL$), number of selected events ($N$),
number of background events ($N_{\rm bkg}$), number of events with
five or more photons ($N_{5\gamma}$, $N_{5\gamma,{\rm bkg}}$), and the
measured $K_L$ nuclear interaction length
in NaI(Tl) ($\lambda_{\rm int}$). The first error
in $\lambda_{\rm int}$ is statistical, the second systematic.
}
\begin{tabular}{ccccccc}
$\sqrt{s}$, GeV & $IL$, nb$^{-1}$ & $N$ & $N_{\rm bkg}$ &
$N_{5\gamma}$ & $N_{5\gamma,{\rm bkg}}$ & $\lambda_{\rm int}$, cm\\ \hline
1.02      & 486  & 30521 & 153 & 23840& 127&$31.4\pm0.4 \pm1.6$ \\
1.04      & 70   & 245   & 3.0 & 183  & 2.1&$42.1\pm8.8 \pm2.1$ \\
1.05      & 264  & 556   & 7.2 & 433  & 5.9&$37.4\pm6.1 \pm1.9$ \\
1.06      & 432  & 656   & 6.6 & 514  & 4.7&$37.5\pm5.9 \pm1.9$ \\
1.07--1.08 & 669  & 516   & 8.9 & 391  & 5.1&$45.6\pm7.3 \pm2.6$ \\
1.09--1.10 & 531  & 206   & 5.8 & 156  & 4.2&$37.1\pm7.6 \pm1.9$ \\
1.11--1.13 & 508  & 118   & 6.9 & 82   & 4.4&$43.3\pm10.3 \pm2.2$ \\
1.14--1.16 & 674  & 101   & 5.9 & 73   & 4.1&$35.7\pm7.9 \pm1.8$ \\ 
1.18--1.21 & 1157 &  76   & 7.9 & 51   & 5.4&$48.3\pm13.8\pm2.4$ \\
1.22--1.29 & 1916 &  83   & 9.8 & 55   & 6.7&$54.8\pm15.7\pm2.8$ \\
1.30--1.38 & 2978 &  49   &10.8 & 35   & 7.7&$45.7\pm21.6\pm2.4$ \\
\hline
\end{tabular}
\end{table}

{\it $K_L$ elastic scattering}.
The $K_L$ elastic scattering leads to increase of the effective
calorimeter thickness and approximately the same increase of
the interaction length. The correction to $\lambda_{\rm int}$
varies from 7.4\% at $E=0.51$ GeV up to 1.9\% at $E=0.69$ GeV. 
The systematic uncertainty of this correction is estimated
by variation of the elastic scattering cross section used in MC 
simulation by 30\% and is found to be 2\%.
 
{\it Beam-generated spurious photons}.
About 10\% of data events contain additional spurious photons 
from the beam background. 
The spurious photons lead to the transition of four-photon events to 
the five-photon class and decrease the measured $\lambda_{\rm int}$ value 
by about 10\%. This effect is taken into account in MC simulation: 
the beam-background events recorded during experiment
with a special random trigger are merged  
with simulated signal events. The $e^+e^-\to \omega\to \pi^0\gamma$ events
are used to test that the simulation reproduces well the photon multiplicity
distribution observed in data.
The systematic uncertainty in the $\lambda_{\rm int}$ due to
spurious photons is estimated to be less than 1\%.

{\it Nonmonochromaticity of $K_L$ mesons}.
Radiative corrections, mainly due to the radiative return
to the $\phi$-meson resonance, lead to the deviation of the $K_L$ 
average energy from $\sqrt{s}/2$. This effect is especially
significant in the range $\sqrt{s}=1.06-1.14$ GeV, in which 
the energies of $K_L$ mesons from the reactions $e^+e^-\to K_SK_L$ and
$e^+e^-\to\phi\gamma\to K_SK_L\gamma$ are already sizably different,
while the peaks from these reactions in the $M_{\rm rec}$ spectra
(Fig.~\ref{fig2}b,c) are not well separated. 
The value of the nuclear interaction length measured in $i^{th}$ energy
interval ($\lambda^{\rm meas}_i$) is related to true 
$\lambda_{{\rm int},j}$ values for intervals with lower energies: 
\begin{equation}
\lambda^{\rm meas}_i=\sum_{j=1}^{i}P_{ij}\lambda_{{\rm int},j},\label{rc}
\end{equation}
where $P_{ij}$ is the probability for $K_L$ produced at beam energy from
$i^{th}$ interval to have (due to initial state radiation) energy in 
$j^{th}$ energy interval. The boundaries of the intervals listed in the first 
column of Table~\ref{tab1} are expanded to provide full coverage
of possible $K_L$ energies, e.g. the value of $\lambda_{\rm int}$ at 
$\sqrt{s}=1.06$ GeV are used in Eq.(\ref{rc}) for the range 1.055--1.065 GeV.
The coefficients $P_{ij}$ are determined from the $K_L$
energy spectra obtained using MC simulation for different $\sqrt{s}$. 
The values of diagonal coefficients $P_{ii}$ are about 0.5 for 
$\sqrt{s}=1.06-1.14$ GeV and then increase to about 0.9 at $\sqrt{s}>1.2$ GeV.
The system of the linear equations (\ref{rc}) is solved to 
determine $\lambda_{{\rm int},j}$.
The systematic uncertainty in $\lambda_{\rm int}$ due to the $K_L$ 
nonmonochromaticity is estimated to be 1\% at
$\sqrt{s}=1.02$ GeV, then increases reaching a maximum of 3\% at
$\sqrt{s}=1.08$ GeV, and then decreases to 1\% at $\sqrt{s}=1.38$ GeV.

The procedure of $\lambda_{\rm int}$ measurement described above is tested
using simulated $e^+e^- \to K_SK_L(\gamma)$ events. 
The found nuclear interaction lengths are consistent with the values 
$\lambda_{\rm int}^{\rm MC}$ used in simulation within the statistical 
uncertainties. The same result is obtained
with $\lambda_{\rm int}^{\rm MC}$ varied by 30\%. 
\begin{table}
\centering
\caption{\label{tSyst} The sources of the systematic uncertainty
on $\lambda_{\rm int}$ (\%) for three $K_L$ energies.}
\begin{tabular}{lccc}
Source                                       & 0.52 GeV & 0.54 GeV & 0.69 GeV \\
\hline
$K_L$ elastic scattering                     & 2.0 & 2.0 & 2.0 \\
$K_L$ nonmonochromaticity                    & 1.0 & 3.0 & 1.0 \\
Spurious photons                             & 1.0 & 1.0 & 1.0 \\
$K_L$ interaction out of NaI(Tl)             & 3.4 & 3.4 & 3.4 \\
Background subtraction                       & 1.5 & 1.5 & 1.5 \\
Detection efficiency                         & 2.2 & 2.3 & 2.6 \\
\hline
Total                                        & 5.0 & 5.8 & 5.2 \\
\end{tabular}
\end{table}
\begin{figure}
\centering
\includegraphics[width=0.5\textwidth]{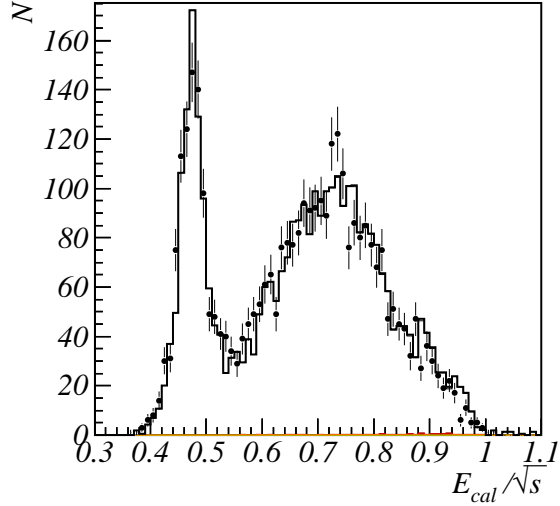}
\caption{\label{fig4}
The distribution of the normalized energy deposition in the
calorimeter for data events (points with error bars) and simulated 
$e^+e^-\to K_SK_L(\gamma)$ events (histogram) at $\sqrt{s}=1.02$ GeV.}
\end{figure}

The corrected values of $\lambda_{\rm int}$ obtained in data 
are listed in Table~\ref{tab1}.
The first error is statistical, the second systematic.
The sources of the systematic uncertainty are listed in Table~\ref{tSyst}.
Some of them are described above. 
The uncertainty due to imperfect simulation of 
$K_L$ nuclear interaction in the detector material before, after, and 
between the calorimeter layers is estimated by variation of
the cross section used in the MC simulation by 30\%.
The uncertainty due to the background subtraction  
is determined by the uncertainties on the numbers of estimated background
events in four- and five-photon classes. The systematic errors
on the detection efficiencies, $\varepsilon^{5\gamma}_i$ and
$\varepsilon^{4\gamma}_i$, associated with $K_S$ reconstruction
are cancelled in the ratio $N_{5\gamma}/(N_{5\gamma}+N_{4\gamma})$.
Remaining systematics is due to imperfect simulation of the 
detector response to $K_L$ mesons.
The distribution of the normalized energy deposition in the calorimeter 
$E_{\rm cal}/\sqrt{s}$ for selected data and simulated events at 
$\sqrt{s}=1.02$ GeV is shown Fig.~\ref{fig4}. Good agreement between data 
and simulation spectra is seen. The narrow peak at 
$E_{\rm cal}/\sqrt{s}\approx 0.5$ corresponds
to events with undetected $K_L$ meson. Since the threshold on the
$K_L$ energy deposition is low (20 MeV), we do not expect any significant 
systematic uncertainty due to inaccuracy in the simulation of 
the detector response. The uncertainty in Table~\ref{tSyst} associated 
with the detection efficiency is determined by simulation statistics.

\section{Conclusion }
\begin{figure}
\centering
\includegraphics[width=0.6\textwidth]{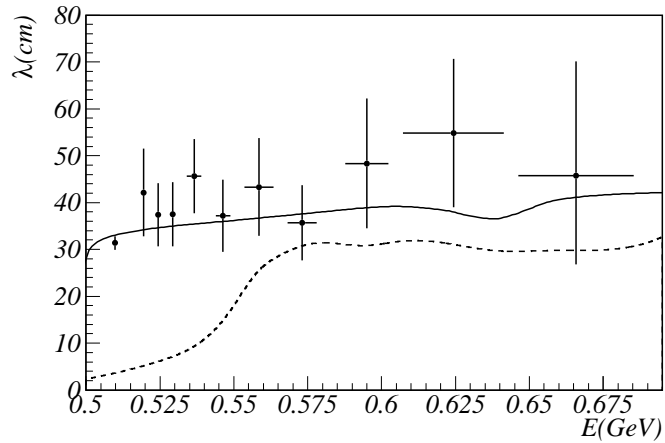}
\caption{\label{fig5} 
The $K_L$ nuclear interaction length in  NaI(Tl) as a function of
the $K_L$ energy. Points with error bars represent the result of this work. 
The solid curve shows the nuclear interaction length used  in the 
UNIMOD~\cite{UNIMOD} simulation package. The dotted line is the same
dependence used in the GEANT4 package, version 9.5 
(low-energy physics)~\cite{GEANT}. }
\end{figure}
  
Using kaons from the reaction $e^+e^-\to K_SK_L$ we have measured the
$K_L$ nuclear interaction length in NaI(Tl) 
in the $K_L$ energy (momentum) range from 0.51 (0.11) 
to 0.69 (0.48) GeV (GeV/$c$).  
The data were collected by the SND detector at the VEPP-2M $e^+e^-$ 
collider. 
We did not find any other measurements of the $K_L$ nuclear length
at such low momenta in literature.

The energy dependence of the $K_L$ nuclear interaction length 
measured in this work is shown in Fig.~\ref{fig5} in comparison
with the nuclear lengths used in the UNIMOD~\cite{UNIMOD} and
GEANT4~\cite{GEANT} packages. 
The nuclear length for GEANT4 is obtained using test simulation,
in which $K_L$ mesons interact in a large block of NaI(Tl).
The $\lambda_{\rm int}^{\rm MC}$ is calculated from the probability
of $K_L$ disappearance in a thin layer of material.

Our data are in good agreement
with UNIMOD. The values of the nuclear length used in GEANT4 
version 9.5 (the model of hadron physics FTFP\string_BERT \cite{GEANTx})
contradict to the measured values at energies below 0.55 GeV. 
The results of this work may be used to refine the  $K_L$
nuclear cross section in GEANT4.

\acknowledgments
This work is supported  by Russian Science Foundation
(project No. 14-50-00080).

\end{document}